\documentclass[fleqn,twoside]{article}

\usepackage{espcrc2}
\usepackage{amsmath,amssymb}
\usepackage[dvips]{graphicx}
\usepackage{array}

\DeclareMathAlphabet{\mathsc}{OT1}{cmr}{m}{sc}

\newcommand{\dof}  {d.o.f.}
\newcommand{\EtAl} {{\it et al.\/}}

\newcommand{\eVq}  {\text{eV}^2}
\newcommand{\Sol}  {\mathsc{sol}}
\newcommand{\Atm}  {\mathsc{atm}}
\newcommand{\JSQ}  {{Just-So$^2$}}

\newcommand{\Dcq}  {\Delta\chi^2}
\newcommand{\Dms}  {\Delta m^2_\Sol}
\newcommand{\Dma}  {\Delta m^2_\Atm}
\newcommand{\Eps}  {\varepsilon}
\newcommand{\Epp}  {\varepsilon'}

\title{Standard and Non-Standard Physics in Neutrino Oscillations}

\author{M.~Maltoni%
  \address{Instituto de F{\'\i}sica Corpuscular -- CSIC/UVEG, \\
    Edificio Institutos de Paterna, Apt.~22085, E-46071 Valencia,
    Spain} 
  } 

\begin{document}

\begin{abstract}
    We analyze the impact of recent solar and atmospheric data in the
    determination of the neutrino oscillation parameters, taking into
    account that both the solar $\nu_e$ and the atmospheric $\nu_\mu$
    may convert to a mixture of active and sterile neutrinos.
    Furthermore, in the context of the atmospheric neutrino problem we
    discuss an extended mechanism of neutrino propagation which
    combines both oscillations and non-standard neutrino-matter
    interactions. We use the most recent neutrino data, including the
    1496-day Super-K solar and atmospheric data samples, the latest
    SNO spectral and day/night solar data, and the final MACRO
    atmospheric results. We confirm the clear preference of all the
    data for pure-active oscillation solutions, bounding the fraction
    of sterile neutrino involved in oscillations to be less than
    $52\%$ in the solar sector and less than $40\%$ in the atmospheric
    sector, at $3\sigma$. For the atmospheric case we also derive a
    bound on the total amount of non-standard neutrino-matter
    interactions, bounding the flavor-changing component to $-0.03
    \leq \Eps \leq 0.02$ and the non-universal component to $|\Epp|
    \leq 0.05$.
\end{abstract}

\maketitle

\section{Introduction}

\begin{table*}[t] \centering\small
    \catcode`?=\active \def?{\hphantom{0}}
    \newcommand{\E}[2]{${#1}\times 10^{#2}$}
    \begin{tabular}{|l|cccc|cccc|}
	\hline
	& \multicolumn{4}{c|}{Active ($\eta_s=0$)} 
	& \multicolumn{4}{c|}{Sterile ($\eta_s=1$)}
	\\
	\hline
	Region
	& $\tan^2\theta_\Sol$ & $\Dms$ & $\chi_\Sol^2$ & GOF
	& $\tan^2\theta_\Sol$ & $\Dms$ & $\chi_\Sol^2$ & GOF
	\\
	\hline
	LMA  & 0.44        & \E{6.6}{-5?} & ?66.1 & 85\% & 0.38        & \E{1.5}{-4?} & ?99.0 & ?6\% \\
	LOW  & 0.66        & \E{7.9}{-8?} & ?75.1 & 60\% & 1.7?        & \E{1.1}{-9?} & 102.0 & ?4\% \\
	VAC  & 1.7?        & \E{6.3}{-10} & ?75.0 & 61\% & 0.19        & \E{2.6}{-10} & ?89.0 & 21\% \\
	SMA  & \E{1.3}{-3} & \E{5.2}{-6?} & ?89.3 & 20\% & \E{3.6}{-4} & \E{3.5}{-6?} & ?99.4 & ?6\% \\
	\JSQ & 1.0?        & \E{5.5}{-12} & ?97.0 & ?8\% & 1.0?        & \E{5.5}{-12} & ?97.5 & ?8\% \\
	\hline
    \end{tabular}
    \caption{ \label{tab:solar}%
      Best fit values of $\Dms$ and $\theta_\Sol$, with the
      corresponding $\chi^2_\Sol$ and the GOF for $81-2$ \dof, for
      pure-active and pure-sterile solar neutrino oscillations.}
\end{table*}

The experimental data on atmospheric neutrinos~\cite{skatm,macroNew}
show, in the muon-type events, a clear deficit which cannot be
accounted for without invoking non-standard neutrino physics. 
In addition to this, the recent results from the Sudbury Neutrino
Observatory (SNO) on neutral current (NC) events~\cite{SNO-sol} have
added more weight to the already robust evidence that an extension of
the Standard Model of particle physics is necessary in the leptonic
sector. 
Altogether, the most popular explanation of both the solar and the
atmospheric neutrino problem is provided by the neutrino oscillation
hypothesis. However, many alternative attempts to account for neutrino
anomalies without oscillations have recently been
proposed~\cite{Pakvasa:2000nt}.
While the results of solar neutrino experiments still admit very good
alternative explanations~\cite{Guzzo:2001mi}, the atmospheric neutrino
anomaly is so well reproduced by the $\nu_\mu \to \nu_\tau$
oscillation hypothesis~\cite{skatm,Gonzalez-Garcia:2002mu} that one
can use the robustness of this interpretation to place stringent
limits on a number of alternative mechanisms~\cite{Fornengo:2001pm}.

In the present work we present an updated analysis of the solar and
atmospheric neutrino data, first in the context of the standard
oscillation hypothesis (Secs.~\ref{sec:solar} and \ref{sec:atmos}),
and then assuming that neutrino posses also non-standard interactions
with matter (Sec.~\ref{sec:nsnmi}).
Motivated by the stringent limits from reactor
experiments~\cite{reactor}, we adopt an effective two-neutrino
approach in which solar and atmospheric analyses decouple. However, in
the pure-oscillation case presented in Secs.~\ref{sec:solar} and
\ref{sec:atmos} our effective two-neutrino oscillation approach is
generalized in the sense that it takes into account that a light
sterile neutrino, advocated to account for the LSND
anomaly~\cite{LSND}, may participate in the conversion
process~\cite{Maltoni:2002ni}.
The natural setting for such a light sterile neutrino is provided by
four-neutrino models. In this paper we will determine the constraints
on the oscillation parameters in this generalized scenario following
from solar and atmospheric data {\it separately}, addressing the
reader to our recent papers in
Refs.~\cite{Maltoni:2001bc,Maltoni:2002xd} for a mass-scheme-dependent
combined analysis of all current oscillation data.

\section{Solar neutrino oscillations}
\label{sec:solar}

In the following we analyze solar neutrino data in the general
framework of mixed active-sterile neutrino oscillations. In this case
the electron neutrino produced in the sun converts into a combination
of an active non-electron neutrino $\nu_a$ (which again is a
combination of $\nu_\mu$ and $\nu_\tau$) and a sterile neutrino
$\nu_s$:
\begin{equation}
    \nu_e \to \sqrt{1-\eta_s}\, \nu_a + \sqrt{\eta_s}\, \nu_s \,.
\end{equation}
The parameter $\eta_s$ with $0\leq \eta_s \leq 1$ describes the fraction
of the sterile neutrino participating in the solar oscillations.
Therefore, the oscillation probabilities depend on the three
parameters $\Dms$, $\theta_\Sol$ and $\eta_s$. 

As experimental data, we use the solar neutrino rates of the chlorine
experiment Homestake~\cite{chlorine}, the most recent result of the
gallium experiments SAGE~\cite{sage} and GALLEX/GNO~\cite{gallex_gno},
as well as the 1496-days Super-Kamiokande data sample~\cite{sksol} in
the form of 44 bins and the latest results from SNO presented in
Ref.~\cite{SNO-sol}, in the form of 34 data bins. Therefore, we have a
total of $3+44+34=81$ observables, which we fit in terms of the three
parameters $\Dms$, $\theta_\Sol$ and $\eta_s$. Further details about
our statistical analysis can be found in~\cite{Maltoni:2002ni}.

\begin{figure*}[t] \centering
    \includegraphics[width=0.98\linewidth]{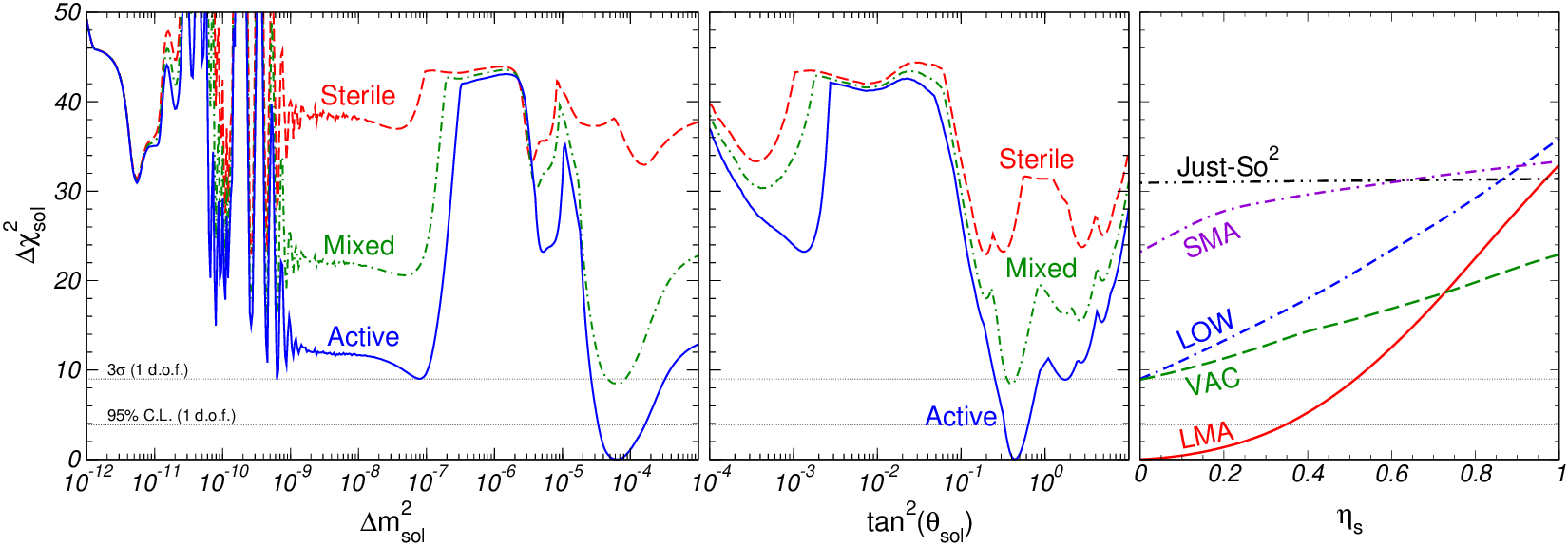}
    \vspace{-9mm}
    \caption{ \label{fig:solar}%
      $\Dcq_\Sol$ as a function of $\Dms$, $\tan^2\theta_\Sol$ and
      $\eta_s$. In each panel the undisplayed parameters are
      integrated out. The ``Active'', ``Mixed'' and ``Sterile'' labels
      correspond to $\eta_s =0$, $0.5$ and $1$, respectively.}
\end{figure*}

We have performed a global fit of solar neutrino data, whose results
are summarized in Tab.~\ref{tab:solar} and Fig.~\ref{fig:solar}. Our
global best fit point occurs in the LMA region, for $\tan^2\theta_\Sol
= 0.44$, $\Dms = 6.6\times 10^{-5}~\eVq$ and $\eta_s=0$. We obtain a
$\chi^2_\mathrm{min}=66.1$, which for $81-3$ \dof\ leads to an
excellent goodness of fit of 83\%. From Fig.~\ref{fig:solar} we can
derive the $3\sigma$ bounds $0.25 \leq \tan^2\theta_\Sol \leq 0.83$,
$2.6\times 10^{-5}~\eVq \leq \Dms \leq 3.3\times 10^{-4}~\eVq$ and
$\eta_s \leq 0.52$. Note that maximal mixing $\theta_\Sol = 45^\circ$
is now ruled out, and that the upper bound on $\Dms$ is rather solid
even without the inclusion of reactor experiments~\cite{reactor}.

From Tab.~\ref{tab:solar} we notice the strong discrimination against
non-LMA solutions implied by the present data: in the pure-active case
we find $\Dcq_\mathsc{low} = 9.0$, $\Dcq_\mathsc{vac} = 8.9$,
$\Dcq_\mathsc{sma} = 23.2$ and $\Dcq_{\text{\JSQ}} = 30.9$ relative to
the global best fit in the LMA region. This shows that the first hints
in favor of a globally preferred LMA oscillation solution, which
followed mainly from the flatness of the Super-K spectra, have now
become a robust result, thanks to the additional data to which SNO has
contributed significantly. Note that especially SMA and \JSQ\ are
highly disfavored with respect to LMA.

The inclusion of a sterile neutrino participating in the oscillation
process is strongly disfavored by the solar data. From the right panel
of Fig.~\ref{fig:solar} we first see how the preferred LMA solution
survives in the presence of a {\it small} sterile component
characterized by $\eta_s$. However, increasing $\eta_s$ leads to a
clear deterioration of all the oscillation solutions. Notice that
there is a crossing between the LMA and VAC solutions, as a result of
which the best pure-sterile description lies in the vacuum regime.
However, in the global analysis sterile oscillations with $\eta_s=1$
are highly disfavored. We find a $\chi^2$-difference between
pure-active and pure-sterile of $32.9$ in the LMA region, which
decreases to $22.9$ if we allow also for VAC. This means that
pure-sterile oscillations are ruled out at $4.8\sigma$ compared to the
active case. 

In summary, we have found that, as long as the admixture of sterile
neutrinos is acceptably small, the LMA is always the best of the
oscillation solutions, establishing its robustness also in our
generalized oscillation scheme.

\section{Atmospheric neutrino oscillations}
\label{sec:atmos}

In our analysis of atmospheric data we make use of the hierarchy
$\Dms\ll\Dma$ and neglect the solar mass
splitting~\cite{Gonzalez-Garcia:2002mu}. Further, in order to comply
with the strong constraints from reactor experiments~\cite{reactor} we
completely decouple the electron neutrino from atmospheric
oscillations~\cite{Gonzalez-Garcia:2001sq}. In the following we
consider a generalized oscillation scheme in which a light sterile
neutrino takes part in the oscillations~\cite{Maltoni:2002ni}. Such a
scenario requires two more parameters in addition to $\theta_\Atm$ and
$\Dma$. We use the parameters $d_\mu$ and $d_s$ already introduced in
Ref.~\cite{Maltoni:2001bc}, and defined in such a way that $(1-d_\mu)$
and $(1-d_s)$ correspond to the fractions of $\nu_\mu$ and $\nu_s$
participating in oscillations with $\Dma$, respectively.  Hence,
pure-active atmospheric oscillations are recovered in the limit
$d_\mu=0$ and $d_s=1$. 

In addition, we also present a ``restricted'' analysis in which the
$\nu_\mu$ is completely constrained to the atmospheric mass states, so
that $d_\mu=0$. In this limit the parameter $d_s$ has a similar
interpretation as $\eta_s$ introduced in the solar case, and $\nu_\mu$
oscillates into a linear combination of $\nu_\tau$ and $\nu_s$:
\begin{equation}
    d_\mu=0 \,:\quad \nu_\mu \to \sqrt{d_s} \,\nu_\tau + \sqrt{1-d_s}
    \,\nu_s \,.
\end{equation}

For the atmospheric data analysis we use all the charged-current data
from the Super Kamiokande~\cite{skatm} and MACRO~\cite{macroNew}
experiments. The Super-Kamiokande data include the $e$-like and
$\mu$-like data samples of sub- and multi-GeV contained events (10
bins in zenith angle), as well as the stopping (5 angular bins) and
through-going (10 angular bins) up-going muon data events. From MACRO
we use the through-going muon sample divided in 10 angular
bins~\cite{macroNew}. Therefore, we have $65$ observables, which we
fit in terms of the four parameters $\Dma$, $\theta_\Atm$, $d_\mu$ and
$d_s$. Further details can be found in Ref.~\cite{Maltoni:2002ni}.

\begin{figure*}[t] \centering
    \includegraphics[width=0.98\linewidth]{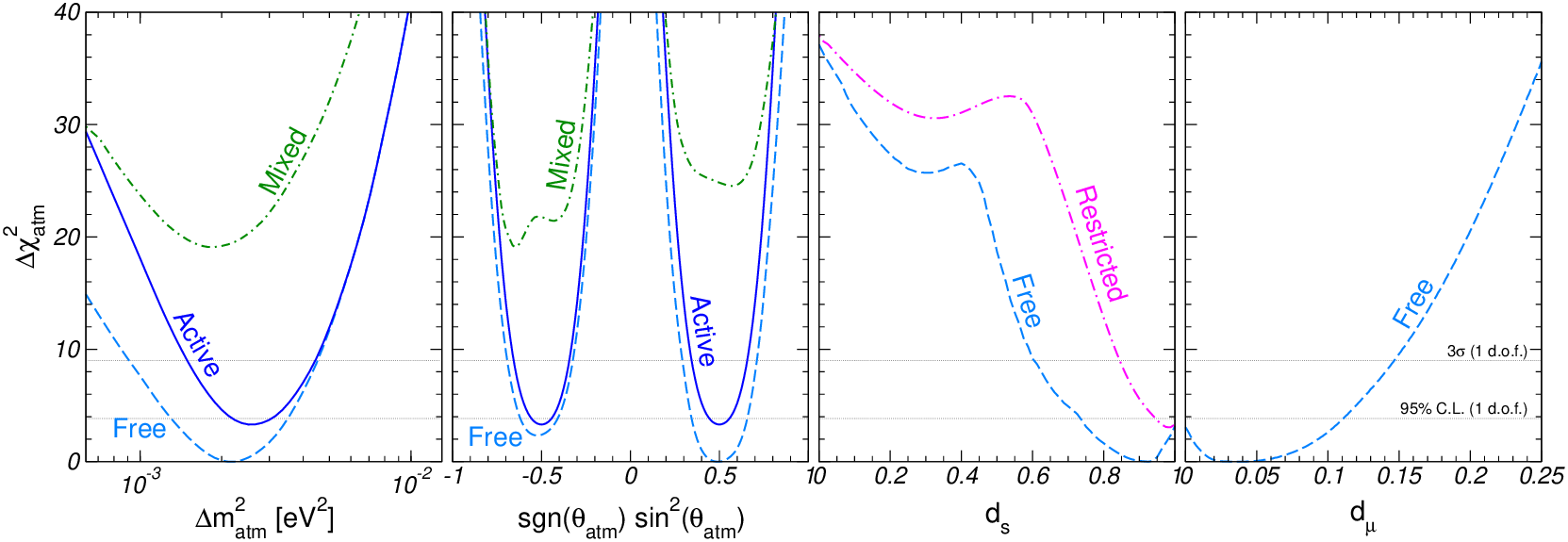}
    \vspace{-9mm}
    \caption{\label{fig:atmos}%
      $\Dcq_\Atm$ as a function of $\Dma$, $\sin^2\theta_\Atm$, $d_s$
      and $d_\mu$. In each panel the undisplayed parameters are
      integrated out. The ``Mixed'' and ``Restricted'' cases
      correspond to $d_s = 0.5$ and $d_\mu = 0$, respectively.}
\end{figure*}

The results of our analysis are summarized in Fig.~\ref{fig:atmos}. In
contrast to the solar case, the atmospheric $\chi^2$ exhibits a
beautiful quadratic behavior, reflecting the high quality of the fit
and the robustness of the oscillation solution.
The global best fit point occurs at $\sin^2\theta_\Atm = 0.49$, $\Dma
= 2.1 \times 10^{-3}~\eVq$, $d_s=0.92$ and $d_\mu=0.04$, and exhibits
a small but non-vanishing sterile neutrino component. However, this
effect is not statistically significant, since also the pure-active
case with $d_s=1$ and $d_\mu=0$ gives an excellent fit: for
$\sin^2\theta_\Atm = 0.5$ and $\Dma = 2.5 \times 10^{-3}~\eVq$, we
have that the $\chi^2$-difference with respect to the best fit point
is only $3.3$.

From Fig.~\ref{fig:atmos} we can extract the $3\sigma$ allowed ranges:
$0.29 \leq \sin^2\theta_\Atm \leq 0.71$ and $9.2 \times 10^{-4}~\eVq
\leq \Dma \leq 4.6 \times 10^{-3}~\eVq$ for the standard two-neutrino
oscillation parameters, while for $d_s$ and $d_\mu$ we have $(1-d_s)
\leq 0.40$ and $(1-d_\mu) \geq 0.85$. So we see that atmospheric data
essentially reject a sterile neutrino component, and strongly bound
the $\nu_\mu$-content in atmospheric oscillations to be nearly
maximal. Note that if we impose the condition $d_\mu=0$, the $3\sigma$
limit on the sterile neutrino fraction improves to $(1-d_s) \leq
0.19$.

These limits on the sterile admixture are significantly stronger than
obtained previously~\cite{Maltoni:2001bc}, and play an important role
in ruling out four-neutrino oscillation solutions in a combined global
analysis of the LSND anomaly~\cite{Maltoni:2002xd}.

\section{Non-standard neutrino interactions}
\label{sec:nsnmi}

\begin{figure*}
    \includegraphics[width=0.98\textwidth]{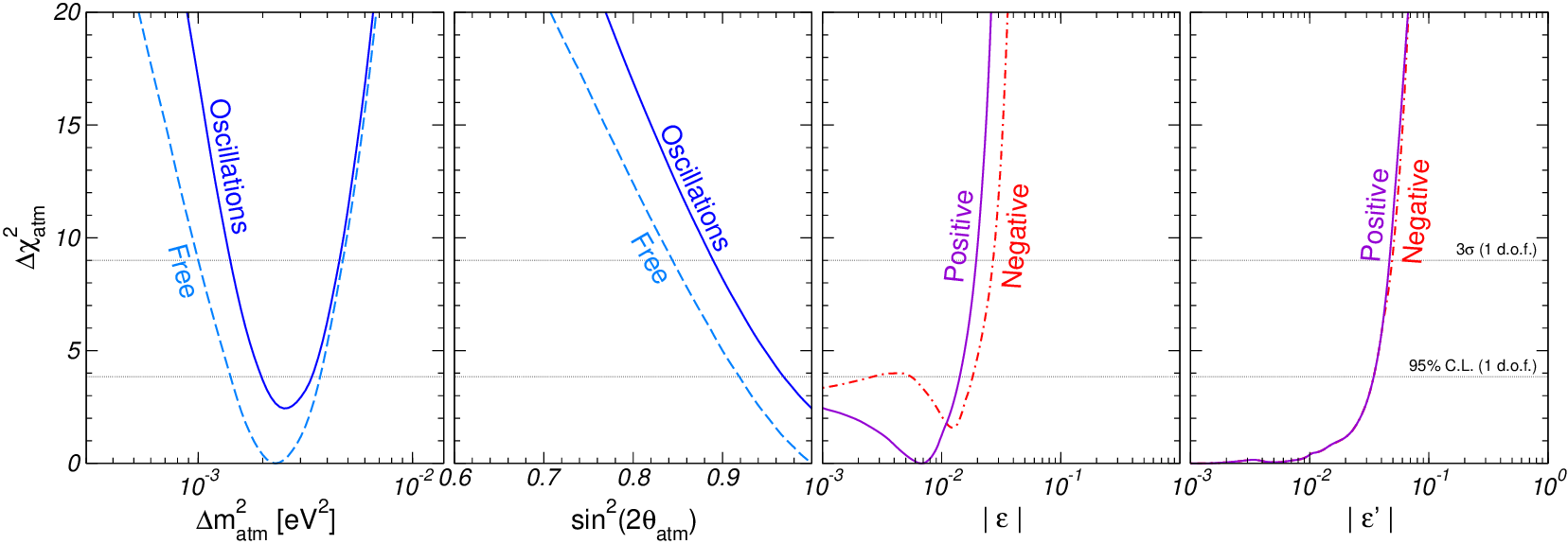}
    \vspace{-9mm}
    \caption{ \label{fig:nsnmi} %
      $\Dcq_\Atm$ as a function of $\Dma$, $\sin^2(2\theta_\Atm)$,
      $|\Eps|$ and $|\Epp|$. The ``Oscillations'' case corresponds to
      $\Eps = \Epp = 0$.  Both ``Positive'' and ``Negative'' values of
      the NSI parameters $\Eps$ and $\Epp$ are displayed.}
\end{figure*}

In this section we still focus on atmospheric data, but instead of
adding a sterile neutrino we now consider the possibility that
neutrinos are massive and moreover possess non-standard interactions
with matter~\cite{Fornengo:2001pm}. This may be regarded as generic in
a large class of theoretical models. For definiteness, in the
following we assume that non-standard neutrino interactions occur only
with the $d$-quark. Also, as in Sec.~\ref{sec:atmos} we completely
decouple $\nu_e$ from atmospheric oscillations. In this case, the
propagation of $\nu_\mu$ and $\nu_\tau$ inside the Earth is governed
by the Hamiltonian:
\begin{equation} \begin{split}
    \mathbf{H} &= \dfrac{\Dma}{4 E}
    \begin{pmatrix}
	-\cos 2\theta_\Atm            & \sin 2\theta_\Atm \\
	\hphantom{-}\sin 2\theta_\Atm & \cos 2\theta_\Atm
    \end{pmatrix} \\
    & \hphantom{=}~ \pm \sqrt{2} \, G_F N_d(r)
    \begin{pmatrix}
	0 & \Eps \\
	\Eps & \Epp
    \end{pmatrix},
\end{split} \end{equation}
where the sign $+$ ($-$) holds for neutrinos (anti-neutrinos),
$N_d(r)$ is the number density of the $d$-quark along the path $r$ of
the neutrinos propagating in the Earth, and $\Eps$ and $\Epp$ are
phenomenological quantities describing flavor-changing and
non-universal non-standard interactions, respectively. Therefore, as
for the sterile neutrino case the transition mechanism depends on four
independent parameters. Note that without loss of generality we can
restrict the range of the oscillation parameters to $0 \leq
\theta_\Atm \leq \pi/4$ and $\Dma \geq 0$, provided that we consider
both positive and negative values of the NSI parameters $\Eps$ and
$\Epp$~\cite{Fornengo:2001pm}. The details of our analysis are the
same as in the previous section.

Our results are summarized in Fig.~\ref{fig:nsnmi}. The global best
fit point occurs at $\sin^2(2\theta_\Atm) = 1$, $\Dma = 2.3\times
10^{-3}~\eVq$, $\Eps = 6.7\times 10^{-3}$ and $\Epp = \pm 1.1\times
10^{-3}$. As in the sterile neutrino case, a small component of NSI is
preferred, however the effect is not statistically significant: the
best pure oscillation solution $\Eps = \Epp = 0$, which occurs at
$\theta_\Atm=45^\circ$ and $\Dma = 2.5\times 10^{-3}~\eVq$, exhibits a
$\chi^2$ which is worse than the global one only by 2.4 units. Thus,
the determination of the oscillation parameters $\Dma$ and
$\theta_\Atm$ is stable under the perturbation introduced by the
additional NSI mechanism, as can be seen also comparing the ``Free''
and ``Oscillations'' lines in the first two panels of
Fig.~\ref{fig:nsnmi}. Moreover, the $\chi^2$ function is quite flat in
the $\Epp$ directions for $\Epp \to 0$, and almost symmetric under the
exchange $\Epp \to -\Epp$.

On the other hand, the pure NSI solution $\Dma = 0$ gives a very poor
fit, and as already found in Ref.~\cite{Fornengo:2001pm} it is
completely ruled out. This occurs since the NSI mechanism is not able
to reconcile the anomaly observed in the upgoing muon sample with that
seen in the contained event sample. Therefore, when combining the two
mechanisms of $\nu_\mu \to \nu_\tau$ transition, oscillations play the
role of leading mechanism, while NSI's can only be present at a
sub-dominant level.

From Fig.~\ref{fig:nsnmi} we also derive the $3\sigma$ allowed ranges:
$0.84 \leq \sin^2(2\theta_\Atm) \leq 1$ and $1.0 \times 10^{-3}~\eVq
\leq \Dma \leq 4.8 \times 10^{-3}~\eVq$ for the oscillation
parameters, while for the NSI parameters we have $-0.03 \leq \Eps \leq
0.02$ and $|\Epp| \leq 0.05$. This is the main result of our analysis,
since it provides limits to non-standard neutrino interactions which
are truly model independent, being obtained from pure neutrino-physics
processes. In particular they do not rely on any relation between
neutrinos and charged lepton interactions. Therefore our bounds are
totally complementary to what may be derived on the basis of
conventional accelerator experiments~\cite{Groom:2000in}.

\section{Conclusions}
\label{sec:concl}

In this paper we have presented an updated analysis of solar and
atmospheric data, both in the context of standard oscillation
hypothesis (generalized to account also for a light sterile neutrino),
and assuming that neutrino posses also non-standard interactions with
matter.

We have found that both solar and atmospheric neutrino data are very
well explained by the simplest hypothesis of oscillations into an
active neutrino, and disfavor the inclusion of an extra sterile state.
In fact, at the $3\sigma$ level the fraction of sterile neutrino
involved in oscillations is limited to be less than $52\%$ in the
solar sector and less than $40\%$ in the atmospheric sector.
Furthermore, atmospheric data also constrain non-standard neutrino
interactions with matter, which are bounded to $-0.03 \leq \Eps \leq
0.02$ for the flavor-changing component and $|\Epp| \leq 0.05$ for
the non-universal component.

In all the considered cases the determination of the oscillation
parameters $\Delta m^2$ and $\theta$ is stable under the inclusion of
exotic physics, such as an extra sterile neutrino or non-standard
neutrino-matter interactions. This fact is an evidence in favor of the
three-neutrino oscillation solution to the solar and atmospheric
neutrino problems. In particular, for what concerns the atmospheric
anomaly we conclude that a maximum mixing angle is a solid result,
which must be incorporated into any acceptable particle physics model,
even in the presence of exotic neutrino interactions.

\section*{Acknowledgments}

I wish to thank my collaborators N.\ Fornengo, T.\ Schwetz, R.\
Tom\`as, M.A.\ T\'ortola and J.W.F.\ Valle.  This work was supported
by Spanish grant BFM2002-00345, by the European Commission RTN network
HPRN-CT-2000-00148, by the European Science Foundation network grant
N.~86 and by the European Union Marie-Curie fellowship
HPMF-CT-2000-01008.


\end{document}